\documentclass[showkeywords,aps,prl,preprint,groupedaddress,twocolumn]{revtex4}
\usepackage{mathtools}

\begin{document}


\title{Von Neumann Entropy from Mean Spin Vector}

\author{Ram Narayan Deb}
\address{Department of Physics, Chandernagore College, Chandernagore, Hooghly, PIN-712136, West Bengal, India}


\email[]{debramnarayan1@gmail.com, debram@rediffmail.com}


\begin{abstract}
We show for a general pure entangled state of two two-level atoms, the von Neumann entropy of the partial traces can be directly measured from the magnitude of the mean spin vector of a single atom of the pair. We emphasize the fact that the von Neumann entropy of the partial traces for such a system can be obtained without knowing the exact form of the quantum state of the two atoms, if we have the value of the magnitude of the mean spin vector of a single atom of the pair. Mean spin vector, used in the context of spin squeezing and spectroscopic squeezing in population spectroscopy, is experimentally measurable and provides an exact measure of von Neumann entropy of the partial traces for such a system. The idea developed in this paper can be used in the context of other quantum mechanical two level systems as the algebra of two level systems can be described by that of  spin-
$\frac{1}{2}$ particles.
\end{abstract}








\maketitle

\section{I. INTRODUCTION}
Quantum entanglement, the basic ingredient of quantum information theory, is an important area of modern days research.  The works of Bennett et al. \cite{Bennett}, Wootters 
\cite{Wootters2, Hill}, Peres \cite{Peres}, Horodecki \cite{Horodecki2}, and several other researchers enriched this field significantly. A measure of quantum entanglement in bipartite states is provided by the von Neumann entropy of the state. In this paper, we show how von Neumann entropy of the partial traces can be obtained for a general pure entangled state of two two-level atoms if we measure the magnitude of the mean spin vector of a single atom of the pair. The magnitude of the mean spin vector is sufficient here to provide the exact information about this entanglement entropy and it is not necessary to know the exact form of the quantum state of the two atoms.

Mean spin vector is an important physical quantity, which has practical importance in the study of spin squeezing and spectroscopic squeezing in the context of population spectroscopy \cite{Wineland2}. We show how the magnitude of this mean spin vector of any single atom of the pair gives us the von Neumann entropy of the partial traces of the system when the exact form of the state is not known.

\section{II. VON NEUMANN ENTROPY OF THE PARTIAL TRACES FROM THE MAGNITUDE OF THE MEAN SPIN VECTOR OF ANY SINGLE ATOM FROM THE 
PAIR OF A                                                                                                                                                                                                                     GENERAL PURE ENTANGLED STATE OF TWO TWO-LEVEL ATOMS}

An atom has many electronic energy levels, but when                                                                                                                                                                                                       it is interacting with a monochromatic electromagnetic radiation, we concentrate only on two of its energy levels among which the transition of the atom takes place. So, we call it as a two                                                                                                                                                 level atom. If the upper and lower energy levels of the n-th atom are $|u_n\rangle$ and 
$|l_n\rangle$ respectively, then, we can construct the following three operators for the atom as ($\hbar = 1$)
\begin{eqnarray}
\hat{J}_{n_x} &=& (1/2)\big(|u_n\rangle\langle l_n| + |l_n\rangle
\langle u_n|\big),\label{1.1a1}\\
\hat{J}_{n_y} &=& (-i/2)\big(|u_n\rangle\langle l_n| - |l_n\rangle
\langle u_n|\big),\label{1.1a2}\\
\hat{J}_{n_z} &=& 
(1/2)\big(|u_n\rangle\langle u_n| - |l_n\rangle\langle l_n|
\big).
\label{1.1a3}
\end{eqnarray}
The operator ${\bf\hat{J}_n}$  is equivalent to a spin-$\frac{1}{2}$ angular momentum operator, since it operates on a two-dimensional complex vector space and since the commutators satisfy the same algebra as
\begin{equation}
 [\hat{J}_{n_x}, \hat{J}_{n_y}] = i\hat{J}_{n_z},
\label{1.2}
\end{equation}
 and two more relations with cyclic changes in $x$, $y$ and $z$ \cite{Itano}.
The third component of ${\bf\hat{J}_n}$ is proportional to the internal energy operator.

Now, we can construct the total angular momentum operator for the two atoms as
\begin{equation}
{\bf\hat{J}} = {\bf\hat{J}_1}\otimes \hat{I} + \hat{I} \otimes {\bf\hat{J}_2},
\end{equation}
where $\hat{I}$ in the first term on the right hand side is the identity operator in the spin-space of atom 2 and $\hat{I}$ in the second term is the identity operator in the spin-space of atom 1. 
The simultaneous eigenvectors of $\hat{J}^2$ and 
$\hat{J}_z$ are denoted as $|j,m\rangle$, such that 
(with $\hbar = 1$)
\begin{equation} 
\hat{J}^2 |j,m\rangle = j(j+1) |j,m\rangle
\end{equation}
and
\begin{equation}
\hat{J}_z |j,m\rangle = m|j,m\rangle,
\end{equation} 
where $m$ takes values $-j, -j+1,...j$.

Now, a general pure entangled state of two two-level atoms is given as
\begin{equation}
|\Psi\rangle = \sum_{i=1}^{2}C_i |u_i\rangle\otimes|v_i\rangle,
\label{1.3}
\end{equation}
where $C_i$, are real numbers (Schmidt coefficients) and 
$|u_i\rangle$ and $|v_i\rangle$ are the quantum state vectors of atoms 1 and 2 respectively \cite{Bennett}.

The von Nemann entropy of the partial traces for the above state is given as \cite{Bennett}
\begin{equation}
S = - {C_1}^2 log_2 {C_1}^2 - {C_2}^2 log_2 {C_2}^2.
\label{1.3a1}
\end{equation}

Now, the general pure entangled state of two two-level atoms, given in Eq. (\ref{1.3}), can be written explicitly in the $\{m_1, m_2\}$ representation as 
\begin{eqnarray}
|\Psi\rangle &=& C_1 \bigg[ \Big( c_3\Big\vert\frac{1}{2}\Big\rangle + c_4 \Big\vert-\frac{1}{2}\Big\rangle\Big)\nonumber\\
 &\otimes& \Big(c_5 \Big\vert\frac{1}{2}\Big\rangle + c_6 \Big\vert-\frac{1}{2}\Big\rangle\Big)\bigg]\nonumber\\
 &+& C_2 \bigg[ \Big( c_7\Big\vert\frac{1}{2}\Big\rangle + c_8 \Big\vert-\frac{1}{2}\Big\rangle\Big)\nonumber\\
 &\otimes& \Big(c_9 \Big\vert\frac{1}{2}\Big\rangle + c_{10} \Big\vert-\frac{1}{2}\Big\rangle\Big)
\bigg].  
\label{1.3a2}
\end{eqnarray}
where
\begin{eqnarray}
|u_1\rangle &=& \Big(c_3\Big\vert\frac{1}{2}\Big\rangle + c_4 \Big\vert-\frac{1}{2}\Big\rangle\Big)
,\\
|u_2\rangle &=& \Big(c_7\Big\vert\frac{1}{2}\Big\rangle + c_8 \Big\vert-\frac{1}{2}\Big\rangle\Big)
\label{1.4}
\end{eqnarray}
are the orthonormal states of atom 1 and
\begin{eqnarray}
|v_1\rangle &=& \Big(c_5\Big\vert\frac{1}{2}\Big\rangle + c_6 \Big\vert-\frac{1}{2}\Big\rangle\Big)
,\\
|v_2\rangle &=& \Big(c_9\Big\vert\frac{1}{2}\Big\rangle + c_{10} \Big\vert-\frac{1}{2}\Big\rangle\Big)
\label{1.4a1}
\end{eqnarray}
are the orthonormal states of atom 2.

Here $C_1, C_2$ are real numbers and $ c_3, c_4, c_5, c_6, c_7, c_8, c_9 $ and $c_{10}$ are in general complex numbers.
Normalization condition on all the above states yields
\begin{eqnarray}
|c_3|^2 + |c_4|^2 &=& 1,\label{1.4a2}\\
|c_5|^2 + |c_6|^2 &=& 1,\label{1.4a3}\\
|c_7|^2 + |c_8|^2 &=& 1,\label{1.4a4}\\
|c_9|^2 + |c_{10}|^2 &=& 1.
\label{1.4a5}
\end{eqnarray}

Using the above equations the normalization condition on $|\Psi\rangle$ yields
\begin{equation}
{C_1}^2 + {C_2}^2 = 1.
\label{1.5}
\end{equation}

Now, the orthogonality condition on the two states of atom 1, $|u_1\rangle$ and $|u_2\rangle$, yields
\begin{equation}
c_3 c_7^{\star} = - c_4 c_8^{\star},
\label{1.6} 
\end{equation}
where $\star$ denotes complex conjugate.
Now, multiplying Eq. (\ref{1.6}) with its complex conjugate, we get
\begin{equation}
|c_3|^2 |c_7|^2 = |c_4|^2 |c_8|^2.
\label{1.7} 
\end{equation}
Using Eq. (\ref{1.4a4}) in the above 
Eq. (\ref{1.7}), we get
\begin{equation}
|c_3|^2 = |c_8|^2.
\label{1.8}
\end{equation}
Similarly, using Eq. (\ref{1.4a2}) in 
Eq. (\ref{1.7}), we get
\begin{equation}
|c_4|^2 = |c_7|^2.
\label{1.9}
\end{equation}
Similarly, the orthogonality condition on the two states of atom 2, $|v_1\rangle$ and $|v_2\rangle$,
yields
\begin{equation}
c_5 c_9^{\star} = - c_6 c_{10}^{\star}.
\label{1.10}
\end{equation}
Using Eq. (\ref{1.4a3}) and (\ref{1.4a5}) in the above Eq. (\ref{1.10}), we obtain
\begin{eqnarray}
|c_6|^2 = |c_9|^2, ~~~~~|c_5|^2 = |c_{10}|^2.
\label{1.11}
\end{eqnarray}
Now, the mean spin vector for atom 1 over the state
$|\Psi\rangle$ is
\begin{equation}
\langle\hat{\overrightarrow{J_1}}\rangle = 
\langle\hat{J}_{1_x}\rangle\hat{i} +
\langle\hat{J}_{1_y}\rangle\hat{j} +
\langle\hat{J}_{1_z}\rangle\hat{k},
\label{1.11a1} 
\end{equation}
where the expectation values are over the state
$|\Psi\rangle$ and $\hat{i}$, $\hat{j}$ and 
$\hat{k}$ are the unit vectors along the positive
$x$, $y$ and $z$ axes respectively.

 The expectation values of the operators
$\hat{J}_{1_x}$, $\hat{J}_{1_y}$ and 
$\hat{J}_{1_z}$ of the atom 1 over the state 
$|\Psi\rangle$ of Eq. (\ref{1.3a2}) are
\begin{eqnarray}
\langle\Psi|\hat{J}_{1_x}|\Psi\rangle &=& 
\frac{1}{2}\Big[{C_1}^2 (c_3 c_4^{\star} + 
c_3^{\star} c_4) \nonumber\\
&+& {C_2}^2 (c_7 c_8^{\star} 
+ c_7^{\star} c_8) \Big],
\label{1.12}\\
\langle\Psi|\hat{J}_{1_y}|\Psi\rangle &=& 
\frac{1}{2i}\Big[{C_1}^2 (c_3^{\star} c_4 - 
c_3 c_4^{\star}) \nonumber\\
&+& {C_2}^2 (c_7^{\star} c_8 
- c_7 c_8^{\star}) \Big],
\label{1.13}\\
\langle\Psi|\hat{J}_{1_z}|\Psi\rangle &=& 
\frac{1}{2}\Big[{C_1}^2 (|c_3|^2 - |c_4|^2)
\nonumber\\
&+& {C_2}^2 (|c_7|^2 - |c_8|^2) \Big].
\label{1.14}
\end{eqnarray}
Using Eqs. (\ref{1.8}) and (\ref{1.9}), the expectation value $\langle\hat{J}_{1_z}\rangle$ in Eq. (\ref{1.14}) can be simplified to
\begin{equation}
\langle\Psi|\hat{J}_{1_z}|\Psi\rangle = 
\frac{1}{2}\Big[{C_1}^2 - {C_2}^2\Big]
\Big[ |c_3|^2 - |c_4|^2\Big].
\label{1.15}
\end{equation}
Now, squaring $\langle\hat{J}_{1_x}\rangle$, given in Eq. (\ref{1.12}), we obtain
\begin{eqnarray}
\langle\hat{J}_{1_x}\rangle^2 &=& \frac{1}{4}
\Big[{C_1}^4\big({c_3^{\star}}^2 c_4^2 + 
2|c_3|^2 |c_4|^2\nonumber\\
 &+& c_3^2 {c_4^{\star}}^2\big) 
+ {C_2}^4 \big({c_7^{\star}}^2 c_8^2 + 
2|c_7|^2|c_8|^2\nonumber\\
 &+& c_7^2 {c_8^{\star}}^2\big)
+ 2{C_1}^2 {C_2}^2 \big(c_3^{\star} c_4 c_7^{\star} c_8\nonumber\\ 
&+& c_3^{\star} c_4 c_7 c_8^{\star}  
+ c_3 c_4^{\star} c_7^{\star} c_8 + c_3 
c_4^{\star} c_7 c_8^{\star}\big)\Big].\nonumber\\
\label{1.16}
\end{eqnarray}
Similarly squaring $\langle\hat{J}_{1_y}\rangle$, by using Eq. (\ref{1.13}), we obtain
\begin{eqnarray}
\langle\hat{J}_{1_y}\rangle^2 &=& -\frac{1}{4}
\Big[{C_1}^4\big({c_3^{\star}}^2 c_4^2 - 
2|c_3|^2 |c_4|^2\nonumber\\
&+& c_3^2 {c_4^{\star}}^2\big) 
+ {C_2}^4 \big({c_7^{\star}}^2 c_8^2 - 
2|c_7|^2|c_8|^2\nonumber\\
 &+& c_7^2 {c_8^{\star}}^2\big) 
+ 2{C_1}^2 {C_2}^2 \big(c_3^{\star} c_4 c_7^{\star} c_8\nonumber\\
 &-& c_3^{\star} c_4 c_7 c_8^{\star}
- c_3 c_4^{\star} c_7^{\star} c_8 + c_3 
c_4^{\star} c_7 c_8^{\star}\big)\Big].\nonumber\\
\label{1.17}
\end{eqnarray}
Using Eqs. (\ref{1.15}), (\ref{1.16}), (\ref{1.17})
and Eqs. (\ref{1.6}), (\ref{1.8}) and (\ref{1.9}), we obtain the magnitude of the mean spin vector of atom 1 as
\begin{eqnarray}
|\langle\hat{\overrightarrow{J_1}}\rangle| &=& \sqrt{\langle\hat{J}_{1_x}\rangle^2 + \langle\hat{J}_{1_y}\rangle^2 + \langle\hat{J}_{1_z}\rangle^2
}\nonumber\\
 &=& \frac{1}{2}\Big( {C_1}^2 - {C_2}^2 \Big).
\label{1.18}
\end{eqnarray} 
Now, using Eq. (\ref{1.5}) in Eq. (\ref{1.18}), we can write
\begin{eqnarray}
{C_1}^2 &=& \frac{1}{2} + 
|\langle\hat{\overrightarrow{J_1}}\rangle|
\label{1.19}\\
{C_2}^2 &=& \frac{1}{2} - 
|\langle\hat{\overrightarrow{J_1}}\rangle|
\label{1.20}
\end{eqnarray}
Therefore, using Eqs. (\ref{1.19}) and 
(\ref{1.20}) in the expresion of von Neumann entropy, given in Eq. (\ref{1.3a1}), we get
\begin{eqnarray}
S &=& - \Big(\frac{1}{2} + 
|\langle\hat{\overrightarrow{J_1}}\rangle|\Big) ~~log_2~~ \Big(\frac{1}{2} + 
|\langle\hat{\overrightarrow{J_1}}\rangle|\Big)
\nonumber\\
&-& \Big(\frac{1}{2} - 
|\langle\hat{\overrightarrow{J_1}}\rangle|\Big)
~~log_2~~\Big(\frac{1}{2} - 
|\langle\hat{\overrightarrow{J_1}}\rangle|\Big)
\label{1.21a1}
\end{eqnarray} 

Now, as long as the state $|\Psi\rangle$, given in Eq. (\ref{1.3a2}), is an entangled state the constants $C_1$ and 
$C_2$ are never zero. Since $|\Psi\rangle$ is normalized, the constants ${C_1}^2$ and ${C_2}^2$
are the probabilities which are less than 1. 
Therefore, from Eq. (\ref{1.18}), we can conclude that the numerical value of $|\langle\hat{\overrightarrow{J_1}}\rangle|$ is less than 
$\frac{1}{2}$. So, in the second term of the right hand side of Eq. (\ref{1.21a1}), the argument of $log_2$ is never zero. Hence, the expression of 
$S$, given in Eq. (\ref{1.21a1}) is mathematically consistent. 

If the measured value of the magnitude of the mean spin vector 
$\langle\hat{\overrightarrow{J_1}}\rangle$ is equal to $\frac{1}{2}$, then either $C_1$ or 
$C_2$ is zero, and we can conclude that the two atoms are unentangled.

Therefore, we have expressed the von Neumann entropy of the partial traces of the general pure entangled state of two two-level atoms in terms of the magnitude of the mean spin  vector of atom 1.

We, now, express the same in terms of the magnitude of the mean spin vector of atom 2.

The expectation values  
$\langle\hat{J_{2_x}}\rangle$, 
$\langle\hat{J_{2_y}}\rangle$ and
$\langle\hat{J_{2_z}}\rangle$, for the atom 2, over the state
$|\Psi\rangle$, given in Eq. (\ref{1.3a2}), are 
\begin{eqnarray}
\langle\Psi|\hat{J}_{2_x}|\Psi\rangle &=& 
\frac{1}{2}\Big[{C_1}^2 (c_5 c_6^{\star} + 
c_5^{\star} c_6) \nonumber\\
&+& {C_2}^2 (c_9 c_{10}^{\star} 
+ c_9^{\star} c_{10}) \Big],
\label{1.21}\\
\langle\Psi|\hat{J}_{2_y}|\Psi\rangle &=& 
\frac{1}{2i}\Big[{C_1}^2 (c_5^{\star} c_6 - 
c_5 c_6^{\star}) \nonumber\\
&+& {C_2}^2 (c_9^{\star} c_{10} 
- c_9 c_{10}^{\star}) \Big],
\label{1.22}\\
\langle\Psi|\hat{J}_{2_z}|\Psi\rangle &=&  
\frac{1}{2}\Big[{C_1}^2 - {C_2}^2\Big]
\Big[ |c_5|^2 - |c_6|^2\Big].\nonumber\\
\label{1.23}
\end{eqnarray}
Using Eqs. (\ref{1.10}), (\ref{1.11}),
(\ref{1.21}), (\ref{1.22}) and (\ref{1.23}), we obtain the magnitude of the mean spin vector of atom 2 as
\begin{eqnarray}
|\langle\hat{\overrightarrow{J_2}}\rangle| &=& \sqrt{\langle\hat{J}_{2_x}\rangle^2 + \langle\hat{J}_{2_y}\rangle^2 + \langle\hat{J}_{2_z}\rangle^2
}\nonumber\\ 
&=& \frac{1}{2}\Big( {C_1}^2 - {C_2}^2 \Big).
\label{1.24}
\end{eqnarray} 
So, we obtain
\begin{eqnarray}
{C_1}^2 &=& \frac{1}{2} + 
|\langle\hat{\overrightarrow{J_2}}\rangle|
\label{1.25}\\
{C_2}^2 &=& \frac{1}{2} - 
|\langle\hat{\overrightarrow{J_2}}\rangle|.
\label{1.26}
\end{eqnarray}
Hence, the von Neumann entropy of the partial traces
can be obtained from Eqs. (\ref{1.3a1}), 
(\ref{1.25}) and (\ref{1.26}) as
\begin{eqnarray}
S &=& - \Big(\frac{1}{2} + 
|\langle\hat{\overrightarrow{J_2}}\rangle|\Big) ~~log_2~~ \Big(\frac{1}{2} + 
|\langle\hat{\overrightarrow{J_2}}\rangle|\Big)
\nonumber\\
&-& \Big(\frac{1}{2} - 
|\langle\hat{\overrightarrow{J_2}}\rangle|\Big)
~~log_2~~\Big(\frac{1}{2} - 
|\langle\hat{\overrightarrow{J_2}}\rangle|\Big)
\label{1.27}
\end{eqnarray} 

Thus, we have expressed the von Neumann entropy of the partial traces in terms of the magnitude of the mean spin vector of atom 2.

Thus, from Eqs. (\ref{1.21a1}) and (\ref{1.27}), we conclude that if we experimentally find out the magnitude of the mean spin vector of any one atom of the entangled pair, we can calculate the von Neumann entropy of the partial traces of the general pure entangled state of two two-level atoms.

We also emphasize the fact that without knowing the exact values of the constants $C_1$ and $C_2$, it is 
possible to calculate the von Neumann entropy of the partial traces of the quantum state by experimentally measuring the 
magnitude of the mean spin vector 
$|\langle\hat{\overrightarrow{J_1}}\rangle|$ or 
$|\langle\hat{\overrightarrow{J_2}}\rangle|$.

Thus, we have connected von Neumann entropy with an experimentally measurable quantity, that is the magnitude of the mean spin vector of any one atom of the entangled pair.  

The idea developed in this paper can be extended to other quantum mechanical two-level systems as the algebra of two-level systems can be described by that of spin $\frac{1}{2}$ particles.




\end{document}